\documentclass[aps,prl,twocolumn,superscriptaddress]{revtex4-1}

\usepackage{graphicx}
\usepackage{subfigure}
\usepackage{natbib}
\usepackage{amsmath,amssymb,amsfonts} 
\usepackage[usenames,dvipsnames]{color}
\usepackage[colorlinks=true]{hyperref}
\usepackage{bm}

\newcommand*{\fullref}[1]{\hyperref[{#1}]{\autoref*{#1} \nameref*{#1}}}

\begin{document}


\newcommand{\blue}[1]{\textcolor{blue}{#1}}

\title{Laser experiment for the study of accretion dynamics of Young Stellar Objects: design and scaling}


\author{G.~Revet}\affiliation{LULI - CNRS, \'Ecole Polytechnique, CEA: Universit\'e Paris-Saclay; UPMC Univ Paris 06: Sorbonne Universit\'es - F-91128 Palaiseau cedex, France}\affiliation{Institute of Applied Physics, 46 Ulyanov Street, 603950 Nizhny Novgorod, Russia}
\author{B.~Khiar} \affiliation{Flash Center for Computational Science, Department of Astronomy \& Astrophysics, The University of Chicago, IL, United States}
\author{J.~B\'eard}\affiliation{LNCMI, UPR 3228, CNRS-UGA-UPS-INSA, 31400 Toulouse, France}
\author{R.~Bonito}\affiliation{INAF (Istituto Nazionale di Astrofisica) - Osservatorio
Astronomico di Palermo, Palermo, Italy}
\author{S.~Orlando}\affiliation{INAF (Istituto Nazionale di Astrofisica) - Osservatorio
Astronomico di Palermo, Palermo, Italy}
\author{M.~V.~Starodubtsev}\affiliation{Institute of Applied Physics, 46 Ulyanov Street, 603950 Nizhny Novgorod, Russia}
\author{A.~Ciardi} \affiliation{LERMA, Observatoire de Paris, PSL Research University, CNRS, Sorbonne University, UPMC Univ. Paris 06, F-75005, Paris, France}
\author{J.~Fuchs}\affiliation{LULI - CNRS, \'Ecole Polytechnique, CEA: Universit\'e Paris-Saclay; UPMC Univ Paris 06: Sorbonne Universit\'es - F-91128 Palaiseau cedex, France}\affiliation{Institute of Applied Physics, 46 Ulyanov Street, 603950 Nizhny Novgorod, Russia}\affiliation{ELI-NP, "Horia Hulubei" National Institute for Physics and Nuclear Engineering, 30 Reactorului Street, RO-077125, Bucharest-Magurele, Romania}

\begin{abstract}
A new experimental set-up designed to investigate the accretion dynamics in newly born stars is presented. It takes advantage of a magnetically collimated stream produced by coupling a laser-generated expanding plasma to a $2\times 10^{5}~{G}\ (20~{T})$ externally applied magnetic field. The stream is used as the accretion column and is launched onto an obstacle target that mimics the stellar surface. This setup has been used to investigate in details the accretion dynamics, as reported in Ref. \cite{Revet2017}. Here, the characteristics of the stream are detailed and a link between the experimental plasma expansion and a 1D adiabatic expansion model is presented. Dimensionless numbers are also calculated in order to characterize the experimental flow and its closeness to the ideal MHD regime. We build a bridge between our experimental plasma dynamics and the one taking place in the Classical T Tauri Stars (CTTSs), and we find that our set-up is representative of a \textit{high plasma $\beta$} CTTS accretion case.

\end{abstract}
\maketitle

\section*{Introduction}\label{Section_intro}

Accretion of matter occurs in a variety of astronomical objects. Examples include black holes in the center of Active Galactic Nuclei (AGNs), pulsars, binary stars such as white dwarfs accreting material from their companion star, and isolated low mass, pre-main-sequence stars.

The accretion in Classical T Tauri Stars (CTTSs) proceeds through matter extracted from the inner edge of an accretion disk which is connected to the star by the star's magnetic field. Accretion takes place in the form of well collimated magnetized plasma columns where matter falls onto the stellar surface at the free fall velocity\citep{Bouvier2003}. Astrophysical observations of such phenomena infer the accretion column to have a density of about $10^{11}-10^{13}~{cm^{-3}}$ \cite{Calvet1998}, a magnetic field of few hundreds of Gauss to kiloGauss \cite{JohnsKrull2007} and a typical free-fall speed of $100-500~{km\ s^{-1}}$. After impact, the matter is shocked and heated up to temperatures of a few MK.

However direct, finely resolved observations of such a process are well beyond present-day observation capabilities. For instance, the \textit{Chandra} telescope has a resolution of $0.2~{AU}$ at a distance of $0.2\times10^{8}~{AU}$, i.e. the nearest star from the sun, Proxima Centauri. It corresponds to a maximum resolution of an object of radius 20 solar radii, while CTTSs have radius of $1 - 2$ solar radii. In this view, laboratory experiments, and especially laser-created plasma experiments, through the high energy density plasmas that they can create and the set of diagnostics that they can use, offer a platform to help understanding accretion plasma dynamics, with both time and space resolution.

Up to now, in the context of accretion shocks, experiments were used to model the impact of a plasma flow onto an obstacle using the so-called “shock-tube” setup as detailed for example by \citet{Cross2016}. This consists in creating a plasma expansion at the rear surface of a target irradiated on its front surface by a high power laser ($I_{Laser}\sim 10^{14}~{W\ cm^{-2}}$). This laser irradiation launches a shock that propagates in the target, comes out at its rear surface and starts an expansion of the target material. This expansion is then guided with the help of a cylindrical tube to finally hit an obstacle at the other edge of the tube. The supersonic plasma flow thus formed propagates with typical speed of $v_{flow}\sim 200~{km\ s^{-1}}$, temperature of $T_{flow}\sim 2\times 10^{4}~{K}\ (\sim 2~{eV})$, and density of $\rho_{flow}\sim 3\times 10^{-2}~{g\ cm^{-3}}$ \cite{Cross2016}. However if that setup presents some clear benefits as a highly collisional plasma flow, necessary for the formation of the shock, it obviously lacks a magnetic field. In addition, the tube edges apply strong constraints on the plasma dynamics especially near the shock region. We note however that the plasma flow generated such a way (i.e. at the rear surface of the laser-irradiate target) exhibits a good scalability with accreting binary systems (the so-called cataclysmic variables) where a white dwarf accretes material coming from its companion. Accretion in cataclysmic variables exhibits more "extreme" parameters than the one of CTTSs: flow density of about $10^{-7.5}~{g\ cm^{-3}}$, flow speed $v_{flow} \sim 5000~{km\ s^{-1}}$, magnetic field strength $B \sim 10-200~{MG}$ and a resulting post shock temperature of $T_{ps} \sim 10^{8}~{K}$ \cite{Busschaert2015, VanBoxSom2018}. We should add that filling the tube with a high Z gas (often used is Xe gas), offer the possibility to study radiative shocks, where radiations start to impact the hydrodynamic behavior \cite{Drake2011,Chaulagain2015}.

Ref. \cite{Young2017} presents an alternative experimental design to investigate accretion, using a laser-created jet, which is geometrically shaped using a conical target. It is found that this setup is scalable to Herbig Ae/Be objects. These young stellar objects (YSOs) are very similar to Classical T Tauri Stars (CTTSs) but within a slightly higher mass range, and presenting a column density $\rho\sim10^{-11}~{g\ cm^{-3}}$, i.e. $\sim 100$ times higher than the CTTSs.

Here we present a new experimental setup which is scalable to CTTS accretion dynamics and uses a magnetically collimated laser-created plasma expansion at the front face of a laser-irradiated target. Results of the experiments were reported in Ref. \citep{Revet2017}. 

\begin{figure*}[htb]
\includegraphics[width=\linewidth]{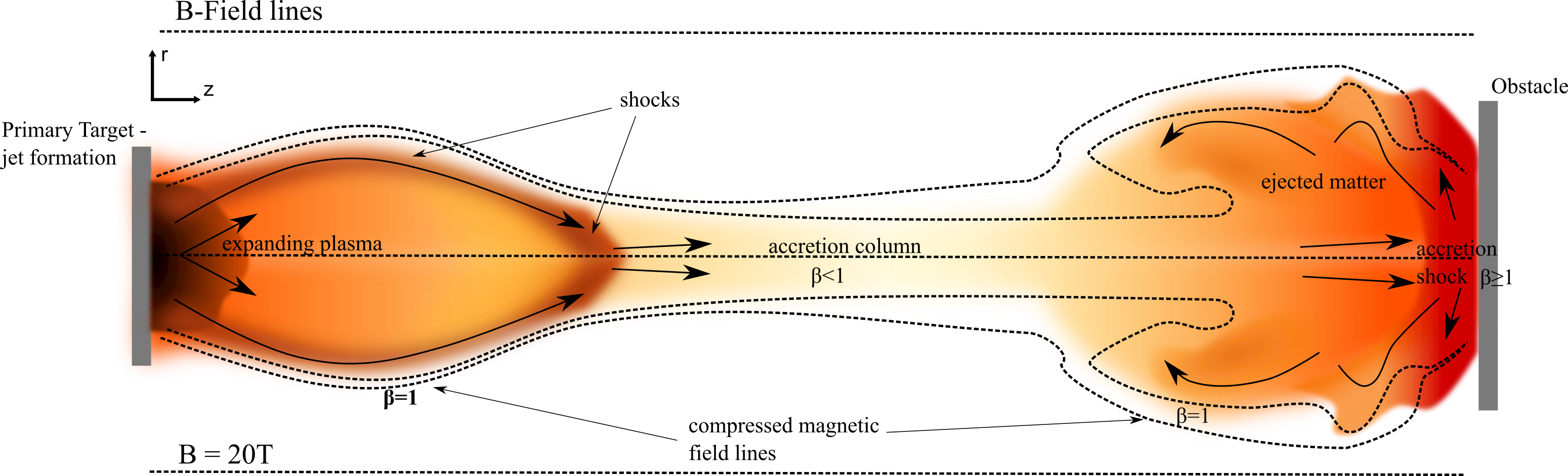}
\caption{Schematic of the accretion experiment performed using a magnetically collimated supersonic flow generated by a laser. The plasma generation and expansion takes place at the primary target location (left side of the image). The jet formed via the interaction with the $2\times 10^{5}~{G}\ (20~{T})$ magnetic field is launched onto a secondary, obstacle target, where the laboratory accretion takes place. As a spatial scale indication, note that the cavity tip is located at $\sim 0.6~{cm}$ from the primary target surface, for a magnetic field of $2\times 10^{5}~{G}\ (20~{T})$ and a laser intensity as the one use at the ELFIE laser facility (see main text). The colors used in the schematic are meant to give indication of the higher density zones (darker) vs. the lower density zones (lighter), the shading of colors used is in no sense quantitative. Ref. \cite{Revet2017} displays results of that accretion setup for a distance between the primary and the obstacle target of about $1.2~{cm}$.
\label{sketch_setup}}
\end{figure*}

\section*{Set-up and plasma flow generation}\label{sec:Section_setup}

The experimental setup consists in using a tens of Joules and one nanosecond duration class laser irradiating a solid target. The plasma exploited for conducting the experiment is the front surface expanding plasma, with the whole dynamic being embedded in an homogeneous externally applied magnetic field, as shown in fig.\ref{sketch_setup}. This accretion shock experimental setup was employed in the work conducted earlier by our group, and published in Ref. \cite{Revet2017}. The experiments were performed on the ELFIE facility (Ecole Polytechnique, France). We used the $60~{J}/0.6~{ns}$ chirped laser pulse, focused on a primary target (PVC material : $C_{2}H_{3}Cl$) onto a $7\times 10^{-2}~{cm}$ diameter focal spot ($I_{max}=1.6\times10^{13}~{W\ cm^{-2}}$) for the plasma expansion creation. The laser-created plasma expanding from the primary target front face was collimated by a $20~{T}$ externally applied magnetic field. The interaction generates a plasma jet with a high aspect ratio (length/radius) which is embedded in the homogeneous magnetic field. The magnetic field is generated by a Helmholtz coil designed to work in a laser environment, a detailed presentation of which can be found in Ref. \cite{Albertazzi2013}.

The mechanisms responsible for the jet collimation is well detailed in Refs. \cite{Albertazzi2014,Ciardi2013,Higginson2017}. It relies on  a pressure balance between the ram pressure of the plasma, $\rho v^{2}$, and the ambient magnetic pressure, $\frac{B^{2}}{2\mu _{0}}$. This pressure balance leads to the formation of a diamagnetic cavity and a curved shock envelope that redirects the plasma flow toward the central axis, where a jet-like flow is finally created. The plasma flow near the laser irradiated target (left) displayed in Fig.\ref{sketch_setup} schematically represents such collimation mechanism.

An extensive description of the characteristics of the jet can be found in Ref. \cite{Higginson2017}. We recall those parameters in Tab.\ref{Comparison_Table}-left column, and hereafter write down the jet main characteristics. The tip propagates at $1000~{km\ s^{-1}}$; that is for the smallest detectable jet tip density ($n_{e}\sim 5\times10^{16}\,-\,10^{17}{cm^{-3}}$ as measured by interferometry at different times), while the $n_{e}\sim 1\times10^{18}{cm^{-3}}$ front propagates at $750~{km\ s^{-1}}$. The electron density then stays quite constant with time and distance.

Regarding the speed and ion density evolution of our plasma flow, an interesting match is found between the experimental flow expansion, and a 1D adiabatic expansion model (detailed below).

\begin{figure}
\includegraphics[width=\linewidth]{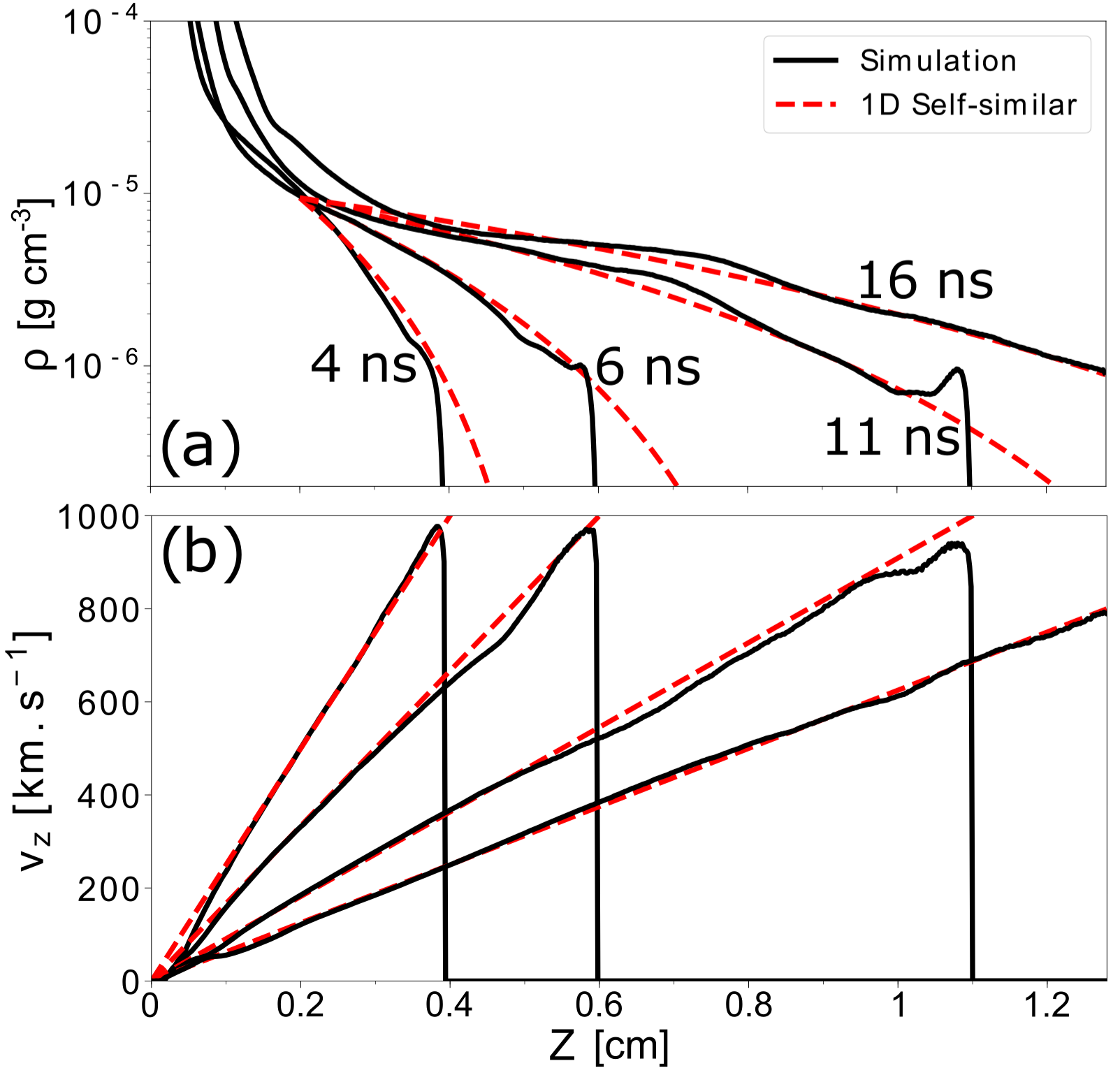}
\caption{GORGON longitudinal density profiles (top) and velocity (bottom) compared to a 1D self-similar analytical model (see text), at different times after the start of the expansion (at $t=0$). The profiles are made via an average around the z-axis over a radius of $7\times 10^{-2}~{cm}$, i.e. the laser focal spot.
\label{Adiabatic_expansion}}
\end{figure}

Indeed, looking at the magnetic field constraint on the plasma in forcing it to flow along a preferential direction (z), essentially reducing an initial 3D expansion to a simple 1D expansion. Fig.\ref{Adiabatic_expansion} represents, with solid black lines, the longitudinal (along z) density and velocity profiles taken from 3D-MHD-resistive simulations of our configuration, using the GORGON code \cite{Ciardi2007,Chittenden2004} and averaged around the z-axis over a radius of $7\times 10^{-2}~{cm}$. The red dashed lines represent a 1D self-similar analytical solution. As one can see, the GORGON results and the 1D solutions match quite well. However, it is necessary to precise that the 1D solution presented here is neither a purely adiabatic solution, nor a purely ballistic expansion. It is actually made of a combination of both approaches. Indeed, while the density profiles are taken from the Landau's self-similar adiabatic solution \cite{L.D.Landau1987} given by:
\begin{equation*}
    \rho=\rho_{0}\left(1-\frac{\gamma-1}{\gamma+1}(1+\frac{z}{c_{s0}t})\right)^{{2}/{(\gamma-1)}}
\end{equation*}
the velocity profiles take their origin within a Lagrangian ballistic solution ($v=z/t$). The Landau solution for the velocity in an adiabatic expansion context would have given instead: $\left|v\right|=\frac{2}{\gamma+1}\left(c_{s0}-\frac{z}{t}\right)$ (i.e., a ballistic solution with an origin moving at the sound speed $c_{s0}$ in the z negative direction).

We stress that the adiabatic solution for the density matches well the simulated data starting from $z=0.2~{cm}$: i.e. z in the adiabatic solution described above, should be replaced by $z+0.2~{cm}$. Also, in order to match the maximum velocity of the experimental/GORGON expansion (i.e. $1000~{km\ s^{-1}}$), the sound speed is artificially increased. A good match is found for $C_{s}^{modified} = 3\times C_{s}$. Finally, as obviously our experimental system is not adiabatic, we call hereafter this one dimensional solution, when referring to it, the \textit{1D self-similar model}.

As shown in Fig.\ref{sketch_setup}, the resulting jet (hereafter called either \textit{stream} or accretion flow) represents the accretion column which then impacts a secondary target ("obstacle") that mimics the stellar surface.
In the experiment of Ref. \cite{Revet2017} this target is made of Teflon material ($CF_{2}$) and it is placed at a distance $z\sim 1.2~{cm}$ from the primary target. At the obstacle location, the impact of the accretion flow generates a reverse shock in the incoming flow. Contrary to the shock-tube setup \cite{Cross2016}, the entire dynamics is here embedded in an external magnetic field, and the edge-free propagation of the flow allows specific plasma motion to freely develop at the border of the reverse shock, as demonstrated in Ref. \cite{Revet2017}.

\section{Dimensionless numbers and plasma parameters}\label{Section_Dimless-numbers}

To understand the astrophysical relevance of the laboratory setup, we now address the scalability between the two systems. Scaling the laboratory flows to astrophysical flows relies on the two systems being described accurately enough by ideal MHD \cite{Ryutov2001, Kurbatov2018}. For the experiment, this generally means to generate a relatively hot, conductive and inviscid plasma, while in the astrophysical case this is often true due to the very large spatial scales involved (see also Table \ref{Comparison_Table} for details about the dimensionless numbers and others plasma parameters). Consequently, the relevant dimensionless parameters, namely the Reynolds number ($R_{e}=L\times v_{stream}/\nu$ ; $L$ the characteristic size of the system ; $v_{stream}$ the flow velocity ; $\nu$ the kinematic viscosity \cite{Ryutov1999}), Peclet number ($P_{e}=L\times v_{stream}/\chi_{th}$ ; $\chi_{th}$ the thermal diffusivity \cite{Ryutov1999}) and Magnetic Reynolds number ($R_{m}=L\times v_{stream}/\chi_{m}$ ; $\chi_{m}$ the magnetic diffusivity \cite{Ryutov2000}) are much greater than one. This ensures the momentum, heat and magnetic diffusion respectively to be negligible with respect to the advective transport of these quantities. In addition to these dimensionless numbers we also consider the acoustic Mach number $M=v_{stream}/c_{s}$, where $c_{s}=\sqrt{\gamma (Z k_{B}T_{e}+k_{B}T_{i})/m_{i}}$ is the sound speed, and the  Alfv\'en Mach number $M_A = v_{stream} /v_A $, where $v_{A}=\sqrt{B^{2}/(\mu_{0} \rho)}$ is the Alfv\'en speed.


The ion mean free path (mfp) should also be smaller than the typical length scale ($L\sim0.1~{cm}$) of the laboratory experiment. Regarding this mean free path, we distinguish between the ion thermal mean free path, $mfp_{i\,th}$, and the ion directed mean free path, $mfp_{i\,dir}$. These two mean free paths are defined from their respective collision rates $\nu_{i\,th}$ and $\nu_{i\,dir}$ \cite{Trubnikov1965}:

\begin{widetext}
\begin{equation}
\nu_{i\,th}=\frac{Z_{s}^{2}Z_{i}^{2}e^{4}}{12(\pi\varepsilon_{0})^{2}}\frac{n_{i}\pi^{\frac{1}{2}}}{m_{i}^{\frac{1}{2}}(kT_{i})^{\frac{3}{2}}}ln\varLambda=4.8\times10^{-8}Z_{s}^{2}Z_{i}^{2}n_{i\,[cm^{-3}]}\mu^{-\frac{1}{2}}T_{i\,[eV]}^{-\frac{3}{2}}ln\Lambda_{i/i}
\label{equation_mfp_therm}
\end{equation}
\begin{equation}
\nu_{i\,dir}=\sum\limits_{s=i}^e \left[\left(1+\frac{m_{i}}{m_{s}}\right)\psi(x^{i/s})\right]\nu_{0}^{i/s}
\label{equation_mfp_directed}
\end{equation}
with
\begin{equation*}
\psi(x^{i/s})=\frac{2}{\sqrt{\pi}}\intop_{0}^{x}t^{1/2}e^{-t}dt\,;\quad x^{i/s}=m_{s}v_{i}^{2}/2kT_{s}
\end{equation*}
and
\begin{equation*}
\nu_{0}^{i/s}=\frac{Z_{s}^{2}Z_{i}^{2}e^{4}}{(4\pi\varepsilon_{0})^{2}}\frac{4\pi n_{s}}{m_{i}^{2}\Delta v^{3}}ln\varLambda=2.4\times10^{-4}Z_{s}^{2}Z_{i}^{2}n_{s\,[cm^{-3}]}\mu^{-2}\Delta v_{\,[km\ s^{-1}]}^{-3}ln\Lambda_{i/s}
\end{equation*}
\end{widetext}

where s = ion or electron: the field particles on which the ion test particle is colliding. $\mu$ is the ion mass in proton mass unit ($\mu=\frac{m_{i}}{m_{p}}$), $e$ the elementary charge, $\varepsilon_{0}$ the vacuum permittivity,  $Z_{i}$ and $Z_{s}$ are the ion charge state of the test and field particles respectively, and $ln\Lambda_{i/s}$ is the Coulomb logarithm \cite{Huba2016}.
We note $\nu_{i\,th}$ to be a limit case of $\nu_{i\,dir}$ for $x^{i/s}\ll 1$, this is to say when the thermal energy of the field particles dominates the directed energy of the test particle.

Finally, $mfp_{th}=v_{r\,th}/\nu_{i\,th}$ with $v_{r\,th}=\sqrt{\frac{2T_{i}}{m_{i}}}$ the relative ion thermal velocity, integrated over a Maxwellian velocity distribution (the factor 2 comes from the reduced mass). The thermal mean free path, measures the distance in between two collisions due to thermal motions. In order for the the fluid description of a plasma to be correct, the thermal mean free path should be much smaller than the characteristic size of the system.
Similarly, $mfp_{dir}=\Delta v/\nu_{i\,dir}$ with $\Delta v$ the relative stream speed, $|\bm{v_{i}}-\bm{v_{s}}|$, between the test ion and the field particles it is colliding in. The directed mean free path accounts for the distance after which a directed momentum (i.e. the stream directed speed) will undergo an isotropization, while colliding with a background of field particles with a Maxwellian velocity distribution.

The initial collision of the stream with the obstacle occurs in reality with an expanding obstacle medium that is ablated from the x-rays generated by the interaction of the laser with the first target. Interferometry and x-rays radiography measurements of the obstacle expansion, at $t=7~{ns}$, exhibits electron density that consists of a very sharp gradient from the solid density (the experimental measurement is limited at $n_{e}\sim 10^{20}~{cm^{-3}}$) to $n_{e}=5\times10^{18}~{cm^{-3}}$ within a distance of $7.5\times 10^{-3}~{cm}$. Following this sharp front, the plasma has a smoother density profiles, consisting of  a decrease of the electron density from $5\times10^{18}~{cm^{-3}}$ to $10^{17}~{cm^{-3}}$ over a distance of $\sim 0.13~{cm}$. One dimensional ESTHER simulations \cite{Colombier2005} matching the experimental expansion indicate a plasma temperature of the obstacle of about $10^{4} - 5\times 10^{4}~{K}$ ($1 - 5~{eV}$, corresponding to an ion charge state of about 1.5 for $CF_{2}$ in the density range $10^{18} - 10^{19} ~{cm^{-3}}$). 
The initial stream collision with the obstacle material is effective at the foot of the sharp density gradient. In this region, an electron density $n_e\sim 6\times10^{18}~{cm^{-3}}$ corresponds to a directed mean free $mfp_{dir} \lesssim 10^{-2}$~cm of the order of the density scale-height; we have used eq. \ref{equation_mfp_directed} with $T_{s}=3~{eV}$, $Z_{i}=2.5$, $Z_{s}=1.5$, and $\Delta v=750~{km\ s^{-1}}$.
From that "stopping point", stream-particles are effectively collisional with the background plasma, the particles loose there directed momentum, and ram pressure is transformed into thermal one at a shock which starts to propagate up the incoming flow.

\begin{figure}
\includegraphics[width=\linewidth]{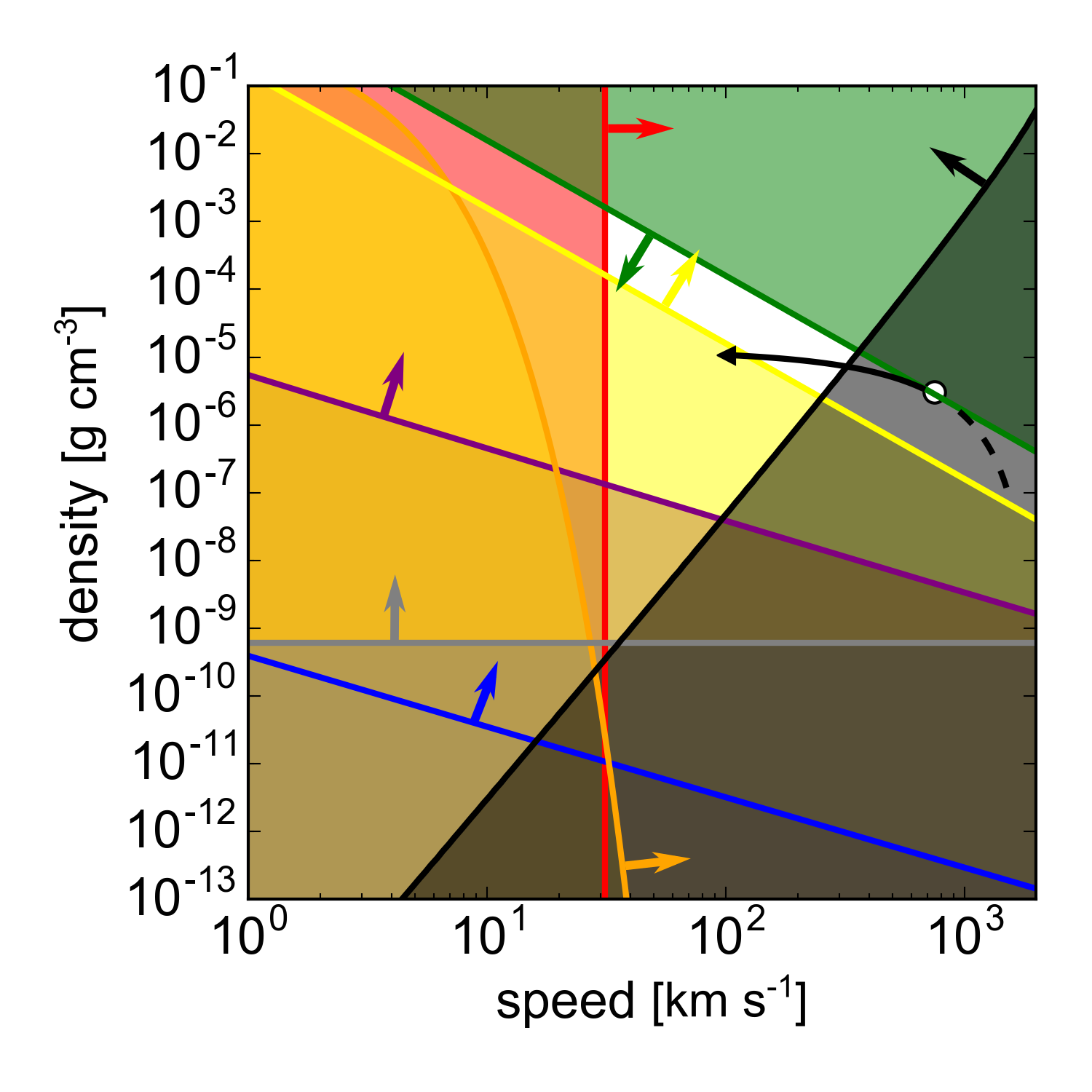}
\caption{Density-speed scalability diagram for the young star accretion experiment \citep{Revet2017}. Each filled-in part of the diagram displays unwanted regions regarding dimensionless numbers. Red : $Mach<1$ ; Blue : $R_{e}<1$ ; Orange : $R_{m}<1$ ; Purple : $P_{e}<1$ ; Gray : $mfp_{ther}>L/10$ ; Black : $mfp_{dir}>L/10$ ; Yellow : $\beta_{dyn}<1$ ; Green : $\beta_{dyn}>10$ . The dimensionless numbers are calculated using the experimental plasma conditions : $L=0.1~{cm}$ ($\sim$ stream radius) ; $T_{e}=T_{i}\sim 10^{5}~{K}\ (10~{eV})$ (except for the $mfp_{dir}$, see main text) ; $A=10.4$ ; $B=2\times 10^{5}~{G}\ (20~{T})$. The white area then represents the area for which the dimensionless numbers respect the scaling constraints. The white point represents the location in the diagram of our initial plasma stream ($v_{stream}=750~{km\ s^{-1}}$ ; $\rho_{stream}\sim3\times10^{-6}~{g\ cm^{-3}}$ - $n_{e}\sim1\times10^{18}~{cm^{-3}}$). The black curve associated to it, materializes the progressive change over time of the stream conditions, following the 1D self-similar expansion of a “reservoir” with density $\rho_{0}=3\times10^{-5}~{g\ cm^{-3}}$, an adiabatic index of $\gamma=\frac{5}{3}$, and an artificially increased initial sound speed (in order to match the plasma maximum expansion speed seen experimentally) - see \fullref{sec:Section_setup} set-up. The dashed curve being the condition before the $v_{stream}=750~{km\ s^{-1}}$ component. To assist the reading of the plot, the arrows anchored to the solid lines indicate the direction for which we obtain the wanted plasma conditions.
\label{v-n_Diagram}}
\end{figure}

While the shock progresses within lower density values of the obstacle medium, and as soon as the density of the obstacle medium becomes sufficiently small compared to the density jump of the compressed stream ($n_{ps}=4\times n_{stream}$ in a strong shock approximation \cite{Zeldovich1966} - the subscript $ps$ stands for \textit{post shock}) it is better to consider the directed mean free path for the collision between the ions of the stream and the ions and the electrons of the stream medium itself, which has already been stopped and shocked. Hence, in eq. \ref{equation_mfp_directed} we take the strong shock condition for the density, i.e. $n_{i}=4\times n_{stream}$, justified by large stream sonic Mach number (see later in the text and Table \ref{Comparison_Table}). For the temperatures, we take $T_{e}\sim T_{stream}\sim 10^{5}~{K}\ (10~{eV})$ and $T_{i}=(\frac{3}{16\times (Z+1)})m_{i}v_{i}^{2}$, expected to be the electrons and ions temperatures just after a strong shock \cite{Zeldovich1966}, and before electron and ion temperature equilibration occurs. We note this temperatures equilibration time to be relatively long and to vary, regarding the density and speed evolution of the stream over time, from about $10$ to $50~{ns}$.

Fig.\ref{v-n_Diagram}, shows a density-speed diagram with filled-in parts representing the unwanted regions regarding the parameters detailed above, while the white area represents the region in the speed-density space for which ideal MHD conditions are satisfied. The white point represents the location in that diagram of our initial plasma conditions. Then, one can see that, while the flow is supersonic ($M_{a}=24\gg1$ -see also Table \ref{Comparison_Table} for a list of plasma parameters), the viscosity can be neglected ($R_{e}=5\times 10^{6}\gg1$), the magnetic field is preferentially advected than dissipated in the plasma ($R_{m}=68\gg1$), the heat advection is dominant over the heat conduction ($P_{e}=7\times10^{2}\gg1$) and that we are in presence of a collisional plasma within the stream itself ($mfp_{ther}\ll L$). This constitutes the necessary conditions for our plasma to be treated in the ideal MHD framework.

\begin{figure}
\includegraphics[width=\linewidth]{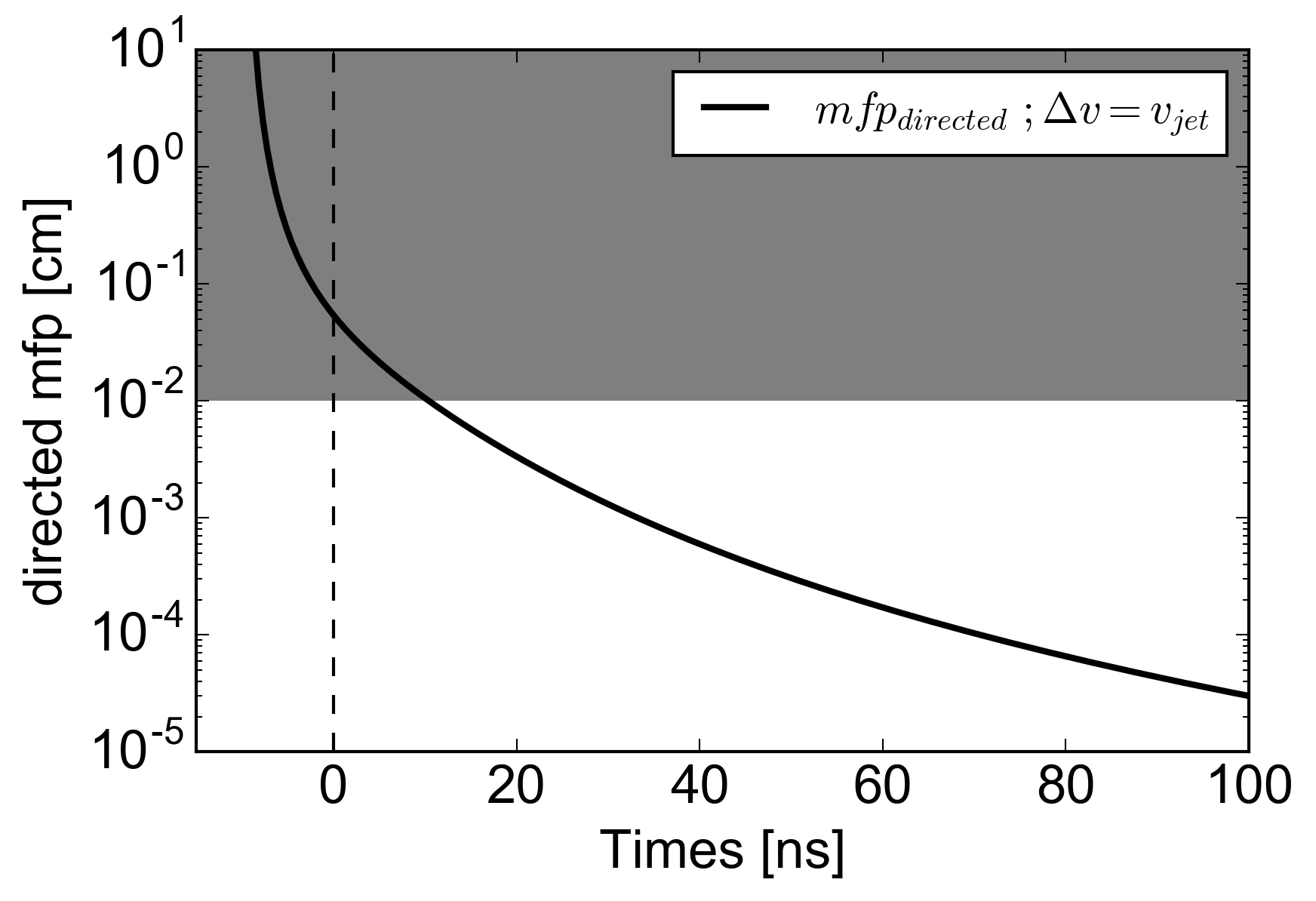}
\caption{Directed mean free path as a function of time, at the obstacle location, i.e. $\sim 1.2~{cm}$ from the stream target, following the 1D self-similar model. The gray area displays the transition region for which $mfp_{dir}=L/10=10^{-2}~{cm}$, where the shock front size over which the particle are stopped becomes sufficiently small compared to the characteristic size of the system $L$, i.e. the stream radius (see also discussion linked to Fig.\ref{v-n_Diagram}). The time in the abscissa takes its origin at the $750~{km\ s^{-1}}$ plasma component arrival time, as also highlighted by the vertical dashed line.
\label{mfp_vs_time}}
\end{figure}

However, we understand the directed mean free path of the flowing stream particles within the shocked previously stream material (black region) to be too large in the initial condition of the stream. Fig.\ref{mfp_vs_time} gives an evolution of that mean free path over time using the 1D self-similar expansion. The time in that figure takes its origin when the $750~{km\ s^{-1}}$ plasma component arrives at the obstacle location. Hence, one can consider that after $10~{ns}$ following the latter component impact, the particles are stopped over a sufficiently small distance ($10^{-2}~{cm}$) in the previously shocked stream material to allow the proper shock formation and propagation. Taking into account that the first collision occurs within the very dense obstacle material, and taking into account the time needed for the shock to propagate and to leave that dense obstacle regions, we could consider the plasma conditions to be, at any time, proper for the shock to form and to propagate away from the obstacle, within the stream material.

\section{Relevance of the experiments to the accretion in Classical T Tauri stars}

As detailed in the introduction, it exists a variety of matter accretion regimes, from Classical T Tauri Stars to Cataclysmic Variables through Herbig Ae/Be objects, regarding the accretion flow density, velocity and magnetic field strength effectively present in these systems.

The laboratory and astrophysical accretion columns are well described by ideal MHD and in order for them to evolve similarly one should verify that the Euler ($Eu=v\sqrt{{\rho}/{P}}$) and Alfv\'en ($Al=B/\sqrt{\mu_{0} P}$) numbers are similar in the two system\cite{Ryutov1999,Ryutov2001}. We define these parameters in the \textit{post-shock region}, where the interaction of the plasma with the magnetic field is largely responsible for determining the accretion shock dynamic regime \cite{Orlando2010}. From the Rankine-Hugoniot relations for a strong shock \cite{Zeldovich1966}, which are valid for hypersonic flows with $M_{a}=v_{stream}/c_{s} \gg1 $  (i.e. $ \rho_{stream} v_{stream}^{2} \gg n_{stream}k_{B}T_{stream}$), the post-shock pressure is given by $P_{ps}=\frac{3}{16\times (Z+1)}\rho _{stream} v_{stream}^{2}$ , $\rho_{ps}=4\times\rho _{stream}$ , $v_{ps}=v_{stream}/4$.  The Euler and Alfv\'en numbers are then given by $Eu=\sqrt{{4(Z+1)}/{3}}$  and $Al=4\sqrt{(Z+1)/3}\times B/(v_{stream}\sqrt{\mu_{0} \rho_{stream}})$. The Euler number only depends on the ion charge state $Z$, which is just a property of the material used in the experiments, while the Alfv\'en number depends on the incoming stream properties and it is proportional to $\beta^{-1/2}_{dyn}$, the plasma dynamic-$\beta$, which is defined as the ratio of the ram to the magnetic pressure, $\beta_{dyn}=\frac{\rho_{stream} v_{stream}^{2}}{{B^{2}}/{2\mu_{0}}}$.
The dynamic-$\beta$ is then the pertinent parameter to look at when trying to link the experimental situation to the astrophysical one, namely the $\beta_{dyn}$ needs to be as close as possible between the two configurations. This is indeed the case since we have $\beta^{exp}_{dyn}=10$ while $\beta^{Chosen\ CTTS}_{dyn}=5$ (see Table \ref{Comparison_Table}).

\begin{figure}
\includegraphics[width=\linewidth]{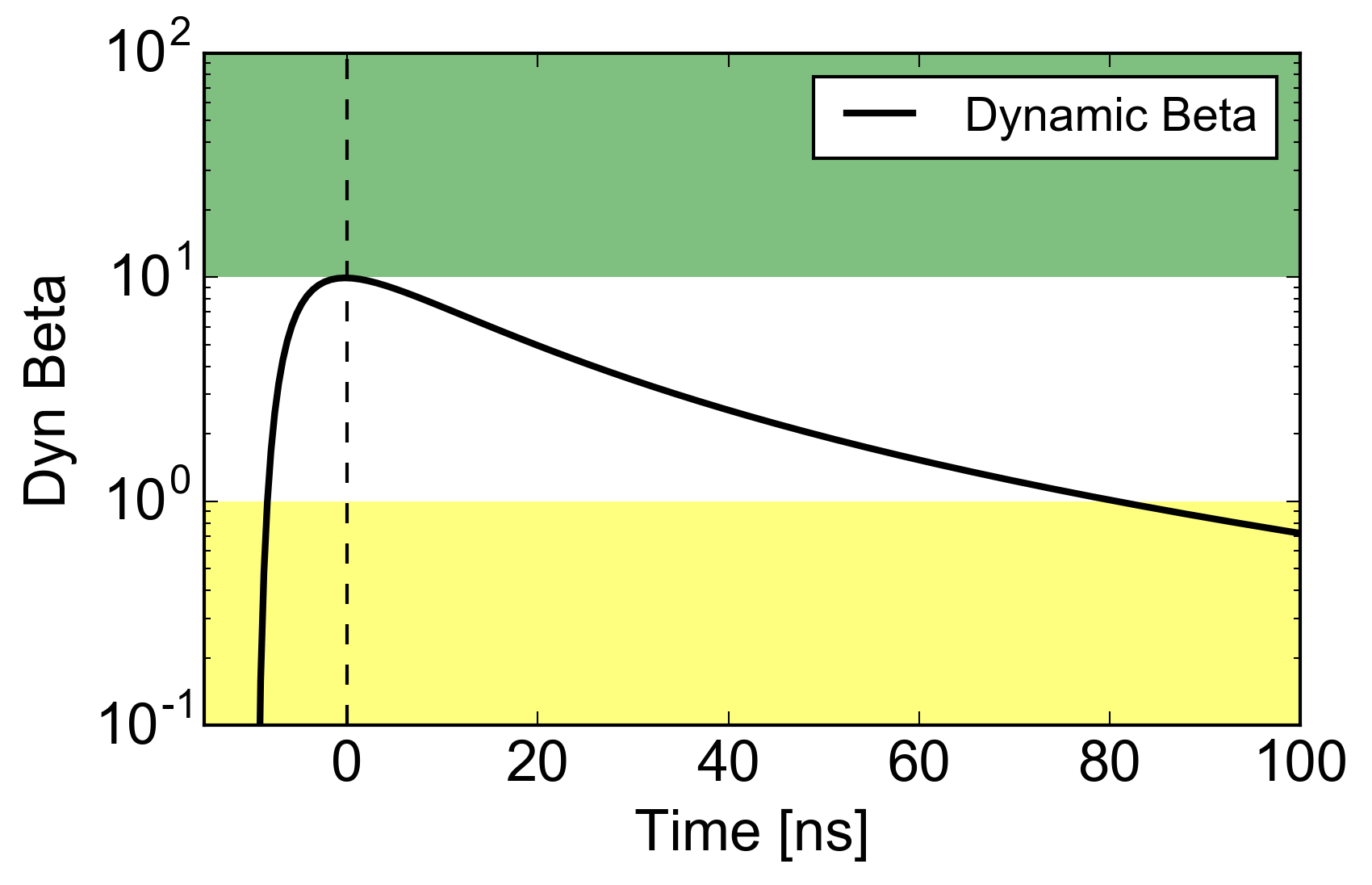}
\caption{Same as Fig.\ref{mfp_vs_time} for $\beta_{dyn}=\frac{\rho_{stream} v_{stream}^{2}}{{B^{2}}/{2\mu_{0}}}$, calculated for $B=2\times 10^{5}~{G}\ (20~{T})$. The green area represents the region $\beta_{dyn} >10$, and the yellow area represents the region $\beta_{dyn} <1$ -keeping the color label of Fig.\ref{v-n_Diagram}. Note that an other interesting information, the ram pressure $\rho_{stream} v_{stream}^{2}$, is directly readable on that plot by multiplying the $\beta_{dyn}$ by the magnetic pressure ${B^{2}}/{2\mu _{0}}= 160~{MPa}$. N.B. Regardless of slight fluctuations it can have within the jet \cite{Ciardi2013}, the initial strength of the magnetic field is the pertinent one in order to characterize the $\beta_{dyn}$, as being the field effectively encountered by the flow at the impact with the obstacle.
\label{betadyn_vs_time}}
\end{figure}

The evolution of $\beta_{dyn}$ (see Fig.\ref{betadyn_vs_time}), following the density and speed evolution given by the 1D self-similar model, indicates that the experimental stream has typical values in the range $\beta_{dyn} \sim 1-10$. In CTTSs, taking standard ion density of about $10^{11}-10^{13}~{cm^{-3}}$ \cite{Calvet1998}, a magnetic field of few hundreds of Gauss to a kiloGauss \cite{JohnsKrull2007} and a typical free-fall speed of $500~{km\ s^{-1}}$, the dynamic-$\beta$ ranges from $\sim 0.01$ to $10$. Which shows that there exists a vast variety of physical conditions in which accretion streams can be found in young stars, and our experiments at $B=2\times 10^{5}~{G}\ (20~{T})$ make it possible to model a high dynamic-$\beta$ (i.e. $\beta_{dyn}>1$)  astrophysical case.
Note that, getting a $\beta_{dyn}=1$ at its maximum, so that the whole accretion dynamic evolves in a magnetic pressure dominated regime, necessitates either to increase the external magnetic field or to decrease the stream expansion speed. The first solution, under the same laser irradiation conditions used in the present study, necessitates a magnetic field strength of $B=6\times 10^{5}~{G}\ (60~{T})$. Such a magnetic field strength could be achievable using the same split Helmholtz coil technology used in the present setup. The second solution necessitates modifying the laser intensity. For this purpose, one can keep in mind the expansion velocity estimate as a function of the laser intensity and laser wavelength: $v^{expansion}_{[cm\ s^{-1}]}=4.6\times10^{7}I_{[10^{14}W\ cm^{-2}]}^{1/3}\lambda_{[\mu m]}^{2/3}$ \cite{Tabak1994}. Conversely, accessing to a higher $\beta_{dyn}$ dynamic will require the use of higher laser intensities or the use of smaller magnetic field strength; both options being easily achievable using the same experimental setup as the one presented in this paper.

\begin{figure}
\includegraphics[width=\linewidth]{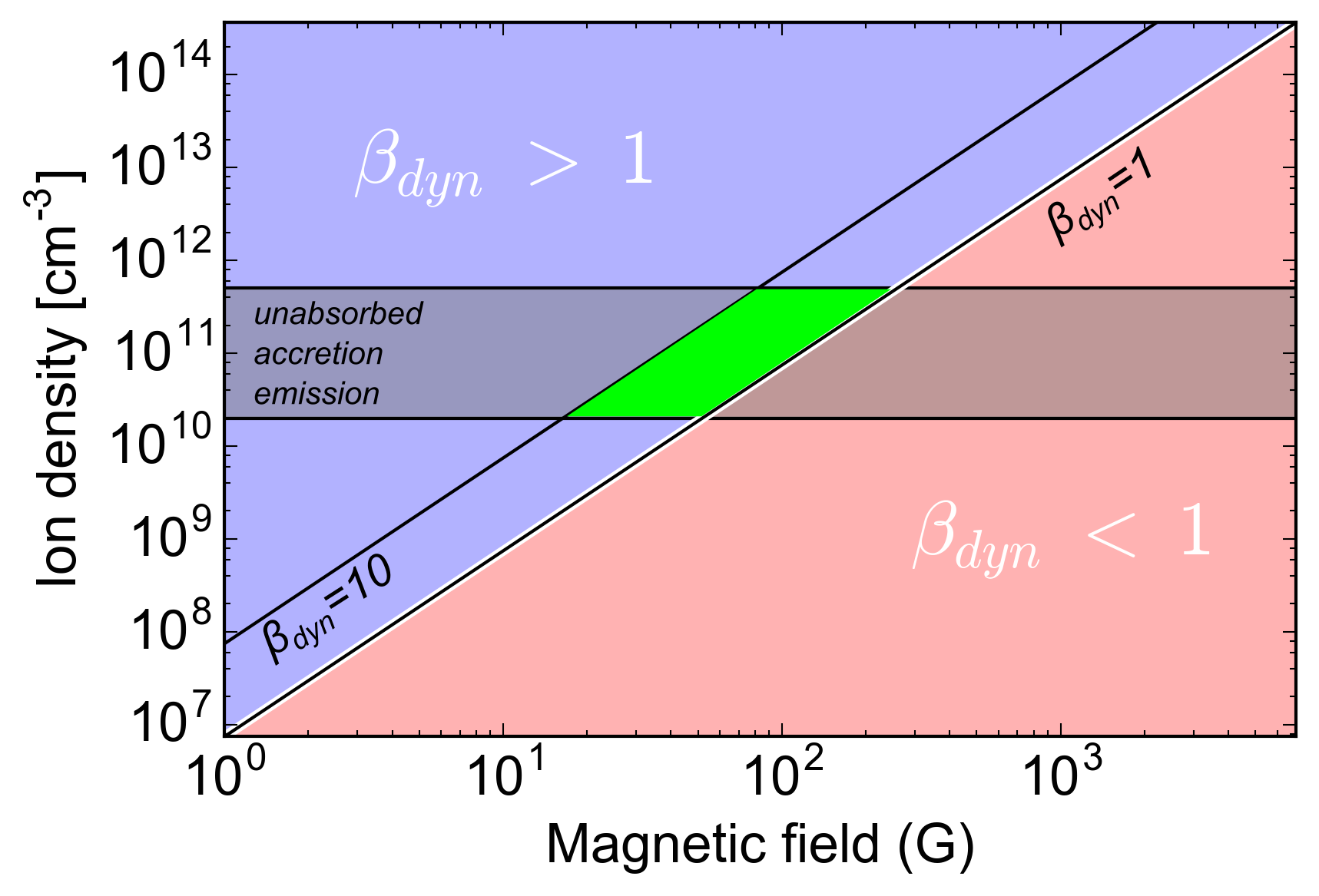}
\caption{Dynamic-$\beta$ variation as a function of the magnetic field and the ion density, for a CTTS accretion column with free fall velocity of $500~{km\ s^{-1}}$. The red filled-in part represents the area for which the couple magnetic field - ion density gives a $\beta_{dyn}<1$, while the blue filled-in zone represents the region for which $\beta_{dyn}>1$. The separation line, $\beta_{dyn}=1$, is represented by the white diagonal. Additional black diagonals display the $\beta_{dyn}=1$ and $\beta_{dyn}=10$ experimental range. The horizontal gray rectangle highlights the $2\times10^{10}$ - $5\times10^{11}~{cm^{-3}}$ density range, corresponding to the observable accretion emission due to the non-absorption of their shocked emissions, through sinking into the chromospheric material (following the study of Ref. \cite{Sacco2010} - see text). The green area represents the CTTS ion density and the magnetic field modeled by the experiment $-$ this leads to an accessible magnetic field strength range of $20-200~{G}$.
\label{B-n_vs_Diagram}}
\end{figure}

Another constraint comes into play when considering a comparison with accreting CTTSs for which accretion radiation emanating from the shocked material is effectively observable. Indeed, as the infalling stream impacts the chromosphere, the exact location for which the stream is halted and the shock starts to develop is where the stream ram pressure is equal to the chromospheric thermal pressure. The ram pressure of the impacting stream can induce important sinking of the stream onto the chromosphere before it to be stopped, if one considers a too large density in the stream. As described in the 1D simulation study of Ref. \cite{Sacco2010}, the shock dynamic can then be buried enough for the accretion emission to be strongly absorbed, and hence hard to detect, which happens for ion stream density above $10^{12}~{cm^{-3}}$. Secondly, an ion stream density below $10^{10}~{cm^{-3}}$ will give a post-shock emissions that cannot be distinguished from coronal emissions. A reasonable density for which the shock dynamic is sufficiently uncovered and distinguishable is thus found to be about $10^{11}~{cm^{-3}}$.

Fig.\ref{B-n_vs_Diagram} represents the ion density as a function of the magnetic field strength for the CTTSs accretion columns. The density restriction discussed above is represented by a gray horizontal rectangle, labeled "unabsorbed accretion emission".
Coupling that density constraint to the $\beta_{dyn}\sim 1-10$ range of the experimental stream (represented by the two black diagonals), one get the green area. It represents the CTTS accretion column parameters the experiment is relevant to model.

This area already gives a good constraint on the astrophysical magnetic field strength that our experimental setup can model : $20~{G}\lesssim B_{CTTS}\lesssim 200~{G}$.

\begin{table}
\begin{tabular}{cc||c}
 & \multicolumn{1}{c||}{\textbf{\textit{Laboratory}}} & \multicolumn{1}{c}{\textbf{\textit{CTTS}}}\tabularnewline
\hline 
\hline 
B-Field $[T]$ & \multicolumn{1}{c||}{$20$} & \multicolumn{1}{c}{$50.10^{-4}$ }\tabularnewline
Material & \multicolumn{1}{c||}{$C_{2}H_{3}Cl$ (PVC)} & \multicolumn{1}{c}{$H$ }\tabularnewline
Atomic number & \multicolumn{1}{c||}{$10.4$} & \multicolumn{1}{c}{$1.28$}\tabularnewline
\hline 
 & \textbf{\textit{Stream}}  & \textbf{\textit{Stream}} \tabularnewline
\hline 
\hline 
Spatial transversal scale $[cm]$ & $0.1$  & $0.5\times10^{10}$\tabularnewline
Charge state & $2.5$  & $1$\tabularnewline
Electron Density $[cm^{-3}]$ & $1\times10^{18}$  & $1\times10^{11}$\tabularnewline
Ion density $[cm^{-3}]$ & $1.9\times10^{17}$  & $1\times10^{11}$\tabularnewline
Density $[g\ cm^{-3}]$ & $3\times10^{-6}$  & $2\times10^{-13}$ \tabularnewline
Ti $[eV]$ & $10$  & $0.22$\tabularnewline
 Flow velocity $[km\  s^{-1}]$ & $100-1000$  & $500$\tabularnewline
Sound speed $[km\  s^{-1}]$ & $\mathrm{31}$ & $\mathrm{7.4}$\tabularnewline
Alfven speed $[km\  s^{-1}]$ & $\mathrm{325}$ & $304$\tabularnewline
Electron mean free path $[cm]$ & $2.7\times10^{-5}$  & $0.7$\tabularnewline
Electron collision time $[ns]$ & $2\times10^{-4}$  & $35$\tabularnewline
Ion mean free path $[cm]$ & $1.4\times10^{-6}$  & $1$\tabularnewline
Ion collision time $[ns]$ & $1.4\times10^{-3}$  & $2.4\times10^{3}$\tabularnewline
Magnetic diffusion time $[ns]$ & $45$  & $4\times10^{21}$\tabularnewline
Electron Larmor radius $[cm]$ & $3.8\times10^{-5}$  & $\mathrm{2\times10^{-2}}$\tabularnewline
Electron gyrofrequency $[s^{-1}]$ & $1.4\times10^{8}$  & $1.2\times10^{8}$\tabularnewline
Ion Larmor radius $[cm]$ & $1\times10^{-3}$  & $\mathrm{1}$\tabularnewline
Ion gyrofrequency $[s^{-1}]$ & $6\times10^{4}$  & $5.2\times10^{4}$\tabularnewline
Electron magnetization & $0.7$  & $30$\tabularnewline
Ion magnetization & $1.4\times10^{-3}$  & $0.9$\tabularnewline
Mach number & $\mathrm{24}$ & $67$\tabularnewline
Alfven Mach number & $\mathrm{2.3}$ & $1.6$\tabularnewline
Reynolds & $5\times10^{6}$  & $6\times10^{11}$\tabularnewline
Magnetic Reynolds & $68$  & $4\times10^{10}$\tabularnewline
Peclet & $7\times10^{2}$  & $6\times10^{9}$\tabularnewline
$\beta_{ther}$ & $1\times10^{-2}$  & $7\times10^{-4}$\tabularnewline
$\beta_{dyn}$ & $10$  & $5$ \tabularnewline
Euler number & $\mathrm{2.9}$  & $1.6$ \tabularnewline
Alfven number & $2.5\times10^{-3}$  & $2.1\times10^{-3}$ \tabularnewline
 & \multicolumn{1}{c}{} & \tabularnewline
\end{tabular}
\caption{Parameters of the laboratory accretion stream, with respect to the ones of the accretion stream in CTTSs, for the incoming stream. The spatial scale corresponds to the stream radius, $0.1~{cm}$ for the laboratory stream while the CTTS accretion column radius is chosen to match the MP Mus infalling radius, retrieved through X-ray measurements of the accretion dynamic \cite{Argiroffi2007}. The stream temperature in the astrophysical case is chosen in order to obtain a stream at thermal equilibrium with the corona. The flow velocity in the laboratory case indicates the full speed range experienced by the stream during its expansion, as described by a 1D self-similar expansion in the \fullref{sec:Section_setup} set-up. The parameters below are however calculated for a stream speed of $750{km\ s^{-1}}$, which is the speed of the $1\times10^{18}{cm^{-3}}$ electron density front.
\label{Comparison_Table}}
\end{table}

For instance, working with a stream density of $n_{stream}=1\times 10^{11}~{cm^{-3}}$, a $\beta _{dyn}\sim5$  will correspond for the astrophysical situation to a magnetic field strength of $\sim50~{G}$.

\bigbreak
\begin{figure}
\includegraphics[width=\linewidth]{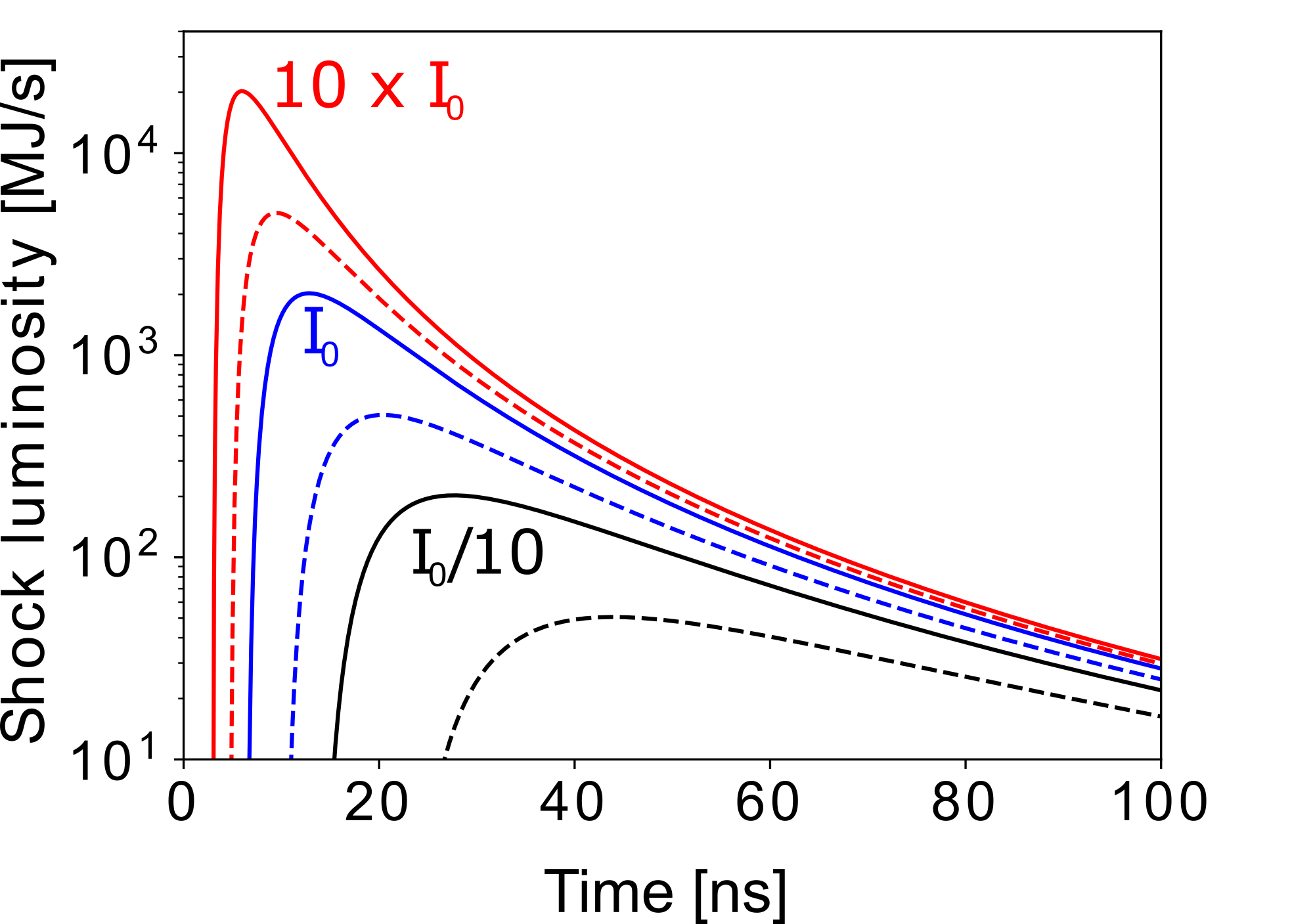}
\caption{Shock luminosity for three different laser intensities (colors) computed at $z=1.2~{cm}$ from the 1D self-similar model described in the main text. Full line: $\lambda=1.06\ \mu m$; dashed line: $\lambda=0.53\ \mu m$. The stream radius is taken to be $0.1~{cm}$.
\label{exp_Lumino}}
\end{figure}

Another important characteristic of an accretion process is the mass accretion rate, $\dot{M}=\frac{dM}{dt}\ [M_\odot\ yr^{-1}]$. The accretion rate is linked to the accretion luminosity $L_{acc}\ [ergs\ s^{-1}]$, which is the luminosity due to the hot continuum excess (i.e. the accretion-produced emission "above" the stellar photospheric emission). Considering the entire directed kinetic energy of the column to be converted into thermal energy, and so into radiation, we have: $L_{acc}=\frac{1}{2}\dot{M}v_{ff}^{2}$ \cite{Gullbring1998}, where $v_{ff}$ is the free fall velocity, i.e. the speed of the accretion flow (from that expression, other corrections as geometrical ones, optical depth or more accurate energy balance considerations can be taken into account). Knowing that $\dot{M}=\frac{dM}{dt}=\rho \times \mathcal{S} \times v_{ff}$, with $\mathcal{S}$ the accretion impact area or equivalently the cross section of the column, we obtain $L_{acc}=\frac{1}{2} \rho \times \mathcal{S} \times v_{ff}^{3}$.
In the experimental context, Fig.\ref{exp_Lumino} displays the \textit{experimental column luminosity} as $L_{acc}^{exp}= \frac{1}{2} \rho_{stream} \times \mathcal{S} \times v_{stream}^{3}$,  for an obstacle target situated at $z=1.2~{cm}$, for different values of laser intensity I (colors) and wavelength $\lambda$ (dashed - solid), and for a stream radius of $0.1~{cm}$. As it can be seen from the 1D self-similar model, the fundamental parameter on which depends the solutions is the initial sound speed of the plasma reservoir (see \fullref{sec:Section_setup} set-up and Ref. \cite{L.D.Landau1987}). By considering an initial steady-state laser ablation in the “deflagration” regime, this sound speed can be expressed as $C_{s} \propto I^{1/3} \lambda^{2/3} A^{-1/3}$ where $I$ and $\lambda$ are the laser intensity and wavelength respectively, and $A$ is the target material mass number. Then, the full blue line in Fig.\ref{exp_Lumino}, corresponding approximately to the parameters given previously in the \fullref{sec:Section_setup} set-up, serves as reference ($I_{0}$) for the others which are obtained by varying $I$ and $\lambda$ using the scaling law for $C_{s}$. The full lines represent solutions at  $\lambda=1.06\ \mu m$ whereas dashed lines represent solutions at  $\lambda=0.53\ \mu m$. One can see that by varying the intensity from $10 \times I_{0}$ to $I_{0}/10$, the luminosity goes from a very picked profile to a relatively flat profile over the typical duration of the experiment. As a result, a high intensity/large wavelength shot would be interesting for studying configurations such as episodic accretion events (we note also the possibility to create a train of streams using multiple laser pulses separated in time, as described in \cite{Higginson2017PRL}).
Oppositely, a low intensity/small wavelength shot should represents a situation closer to the steady accretion configuration usually investigated in astrophysical studies (see Ref. \cite{Orlando2010}).
The luminosity, and so the mass accretion rate being a privileged observable in the astrophysical context, notifying strong experimental plasma dynamic differences at the accretion shock location, linked to different luminosity profiles, makes such a point of interest for parametric studies of the accretion dynamic in the laboratory with a direct and strong anchor in the astrophysical context.

\section*{Conclusion}

We have presented and discussed in this paper a new experimental set-up to recreate in the laboratory magnetized accretion dynamics scalable to Classical T Tauri Stars. The front-surface-target plasma expansion, generated via a laser-solid interaction (tens of Joules / nanosecond duration), is exploited and coupled to an externally applied magnetic field of strength $B=2\times 10^{5}~{G}\ (20~{T})$. Such a coupling generates a collimated jet, the density and velocity of which follows a 1D self-similar expansion. This jet mimics the accretion column, and in order to generate an accretion shock it is launched onto a secondary obstacle that represents the stellar surface. The stream characteristics, at the very beginning of the impact, can be resumed at the impact location as $\rho \sim 3\times10^{-6}~{g\ cm^{-3}}$, $v \sim 750~{km\ s^{-1}}$ and $T_{e}=T_{i} \sim 10^{5}~{K}\ (10~{eV})$. The experimental accretion is shown to be scalable to a CTTS accretion with parameters that are $\rho \sim 10^{-13}~{g\ cm^{-3}}$, $v \sim 500~{km\ s^{-1}}$, $T_{e}=T_{i} \sim 2500~{K}\ (0.22~{eV})$ and $B=20-200~{G}\ (2\times 10^{-3}-2\times 10^{-2}~{T})$. This is to say, a \textit{high dynamic $\beta$} accretion case, compared to what is expected for the \textit{standard} magnetic field strength in CTTSs.
Such a high plasma $\beta$ experimental dynamic exhibits interesting accretion-column-edge-features, as the formation of a plasma cocoon that surrounds the shocked region. The latter could be an explanation for X-rays absorption effects of interest in order to interpret astrophysical observations of those phenomena \cite{Bonito2014}. A description of the results of the experiment, the set-up of which is explained in the present paper, can be found in Ref. \cite{Revet2017}.

\bigbreak
\begin{acknowledgments}
We thank the LULI and LNCMI teams for technical support, B. Albertazzi and M. Nakatsutsumi for their prior work in laying the groundwork for the experimental platform. This work was supported by ANR Blanc Grant n◦ 12-BS09-025-01 SIL- AMPA (France) and by the Ministry of Education and Science of the Russian Federation under Contract No. 14.Z50.31.0007. This work was partly done within the LABEX Plas@Par, the DIM ACAV funded by the Region Ile-de-France, and supported by Grant No. 11-IDEX- 0004-02 from ANR (France). Part of the experimental system is covered by a patent (n◦ 1000183285, 2013, INPI-France). The research leading to these results is supported by Extreme Light Infrastructure Nuclear Physics (ELI-NP) Phase I, a project co-financed by the Romanian Government and European Union through the European Regional Development Fund.
\end{acknowledgments}
\bibliography{ref_HEDLA_paper}

\end{document}